\documentclass[12pt]{article}
\usepackage{arxiv}
\usepackage[utf8]{inputenc} % allow utf-8 input
\usepackage[T1]{fontenc}    % use 8-bit T1 fonts
\usepackage[square,sort,numbers]{natbib}
\usepackage{varioref}
\usepackage[dvipsnames]{xcolor}
\usepackage[colorlinks]{hyperref}       % hyperlinks
\hypersetup{
linkcolor=BrickRed
,citecolor=Green
,filecolor=Mulberry
,urlcolor=NavyBlue
,menucolor=BrickRed
,runcolor=Mulberry
,linkbordercolor=BrickRed
,citebordercolor=Green
,filebordercolor=Mulberry
,urlbordercolor=NavyBlue
,menubordercolor=BrickRed
,runbordercolor=Mulberry
}%for the sake of ease for eyes
\usepackage{caption}
\usepackage{subcaption}
\usepackage{comment}
\usepackage[shortlabels]{enumitem}
\usepackage{url}            % simple URL typesetting
\usepackage{booktabs}       % professional-quality tables
\usepackage{amsfonts}       % blackboard math symbols
\usepackage{nicefrac}       % compact symbols for 1/2, etc.
\usepackage{microtype}      % microtypography
\usepackage{lipsum}		% Can be removed after putting your text content
\usepackage{graphicx}
\usepackage{natbib}
\usepackage{doi}
\usepackage{multirow}
\usepackage{amsmath}

\numberwithin{figure}{section}
\numberwithin{equation}{section}
\usepackage{amssymb}
\usepackage{tikz}
\usetikzlibrary{shapes.geometric,arrows}
\usetikzlibrary{calc}
\usetikzlibrary{positioning}
\usepackage[linesnumbered,ruled,vlined,algosection]{algorithm2e}
\usepackage{setspace}
\usepackage{indentfirst}
\tikzstyle{startstop} = [rectangle, rounded corners,  minimum height=1cm,text centered, draw=black]
\tikzstyle{io} = [trapezium, trapezium left angle=70, trapezium right angle=110, text centered, draw=black]
\tikzstyle{process} = [rectangle, minimum height=1cm, text centered, draw=black]
\tikzstyle{decision} = [diamond,  minimum height=1cm, aspect=2,text centered, draw=black]
\tikzstyle{arrow} = [thick,->,>=stealth]
\usepackage{amsmath}
\DeclareMathOperator*{\argmax}{arg\,max}
\DeclareMathOperator*{\argmin}{arg\,min}
\newcommand*{\Break}{\textbf{break}}

\title{ChemiRise: a data-driven retrosynthesis engine}

\date{\today}	% Here you can change the date presented in the paper title
\author{ 
	Xiangyan Sun, Ke Liu, Yuquan Lin, Lingjie Wu, Haoming Xing, Minghong Gao, \\ \textbf{Ji Liu, Suocheng Tan, Zekun Ni, Qi Han, Junqiu Wu, Jie Fan} \thanks{Corresponding author. jiefan@accutarbio.com} \\
	Accutar Biotechnology Inc.\\
}

\renewcommand{\shorttitle}{\textit{ChemiRise paper}}

\hypersetup{
pdftitle={ChemiRise Paper},
pdfsubject={Machine Learning, Computational Chemistry, Retrosynthesis},
pdfauthor={AAA},
pdfkeywords={},
}

\begin{document}
\maketitle
\begin{abstract}
We have developed an end-to-end, retrosynthesis system, named ChemiRise, that can propose complete retrosynthesis routes for organic compounds rapidly and reliably. The system was trained on a processed patent database of over 3 million organic reactions. Experimental reactions were atom-mapped, clustered, and extracted into reaction templates. We then trained a graph convolutional neural network-based one-step reaction proposer using template embeddings and developed a guiding algorithm on the directed acyclic graph (DAG) of chemical compounds to find the best candidate to explore. The atom-mapping algorithm and the one-step reaction proposer were benchmarked against previous studies and showed better results. The final product was demonstrated by retrosynthesis routes reviewed and rated by human experts, showing satisfying functionality and a potential productivity boost in real-life use cases.
\end{abstract}

% keywords can be removed

\section{Introduction} \label{sec:intro}
The art of organic synthesis is arguably the most challenging skill to master in chemical sciences, due to the intimidating number of published reactions with empirical and often ambiguous rules of application, together with the complexity that originates from the fragility and conflicts of reactivities of various functional groups. At the same time, the expected throughput of organic chemists has increased as synthesis has become a bottleneck of the rational designing process of new, computer-designed molecules. %(TODO: other points that might worth mentioning: increasing synthetic complexity, repetitive work that can be streamlined, etc)

The first productivity boost of organic synthesis transpired after E.J.Corey famously systematized the retrosynthesis procedure, a practice proposed in the 1950s that soon became a norm in organic chemistry~\cite{corey1991retrosynth}. He then pioneered the computerized retrosynthesis technique in 1971 by developing LHASA, a template-based retrosynthesis system~\cite{corey1971centenary,corey1980lhasa1,corey1985lhasa2,pensak1977lhasa}.
Nonetheless, despite gradual improvements that could be attributed primarily to the increase of computing power and available data, the envisaged expert-level retrosynthesis system was not achieved In contrast, many other conundrums have been conquered during the same period, such as GO~\cite{silver2017mastering} and protein folding~\cite{senior2020improved}, with breakthroughs in methodology and a leap forward in performance.

The essence of computer-aid synthetic planning is still based on the retrosynthesis philosophy and most retrosynthesis systems can be divided into three integral components:
\begin{itemize}
	\item A curated knowledge base of organic chemical reactions. Reactions need to be atom mapped, which means assigning the correspondence between atoms in the reactants and the products. Some methods further extract and cluster the reaction cores, i.e. atoms whose bonding status changed during the reaction, into reaction templates.
	\item Single-step retrosynthesis, which proposes the most likely precursors for a compound. Sub-problems in this part include finding applicable reactions, or proper bond-braking, and ranking all proposed reactions.
	\item Guiding algorithms, which evaluate current synthesis routes and determine the next compound to perform the single-step retrosynthesis until a route starting with commercially available compounds is found.
\end{itemize}  

Currently, many published results on this problem have focused on the one-step precursor prediction problem, while few holistic solutions have been developed to meet the demand of real-life, end-to-end uses.

Synthia (previously Chematica), a commercially-available platform is a classic expert system that has over 100,000 hand-coded reaction rules. Numerous heuristic functions are needed to effectively search the hypergraph of reactants and reactions to prune chemically unrealistic possibilities. After over a decade of efforts on hardwiring a gigantic knowledge base in the system, it was reported to deliver impressive yet anecdotal routes verified by laboratory works, with no systematic benchmark tested on it~\cite{klucznik2018efficient}.

To circumvent the consuming work of hand-coding rules which might soon become incomplete and outdated owing to the exponential growth of newly published literature, many researchers preferred automatic information retrieval of existing reactions. ~\cite{coley2017computer} proposed a similarity-based method to evaluate applicable reactions that were successfully performed on molecules that have a structural resemblance to the target molecule, in terms of the Tversky similarity of the Morgan fingerprints of molecules. This method, being template-free, can apply any properly atom mapped reactions without the extraction and clustering of reaction cores. But this similarity-only feature may ignore reactivity conflicts of different functional groups and sacrifice the transferability of some chemical reactions that can be applied to the target compound due to the limited size of the knowledge base.

Inspired by the phenomenal success of machine learning (ML) with neural networks (NN), endeavors have recently been redirected to utilizing architectures that succeeded in other fields. Convenient textual notations of compounds and reactions such as SMILES~\cite{weininger1988smiles} and SMIRKS~\cite{smirk} reduce the dimensionality of the representation and make chemical reactions resemble translation processes, for which various existing NN-based methods have been developed. From seq-to-seq to transfomer models, avant-garde natural language processing (NLP) models are ported to the retrosynthesis problem promptly. Current best performers include the transformer model by Tetkos et al. ~\cite{tetko2020transformer} and the seq2seq model by Liu et al. ~\cite{liu2017retrosynthetic}. The prediction of sequence-based methods, which require heavy training and parametrization, are criticized for being uninterpretable and for the loss of chemical reasoning.~\cite{dai2020retrosynthesis} 

Besides fingerprints and textual notations, graphs can serve as a natural representation of molecules, and features can then be extracted via graph neural networks (GNNs). Some implements of GNNs in the singe-step retrosynthesis problem are reported to surpass the performance of both the expert system and seq2seq model-based methods.~\cite{dai2020retrosynthesis}. 

Although studies that were focused on single-step predictions usually remarked that the final end-to-end solution would merely be a repetitive process of calling the single-step prediction function, it turned out the expansion and search strategy used is beyond such naivete and crucial to a feasible solution, as the branching factor is too large to perform an uncontrolled search in the reaction space~\cite{szymkuc2016computer}. The aforementioned Chematica system claimed to effectively search the Network of Chemistry (NOC) by combining a cost function-based breadth-first-search (BFS) and precalculated, optimized synthetic routes of popular compounds in their data structure.~\cite{szymkuc2016computer} However, such heuristic is arbitrary and not data-driven, and is also expensive to maintain and update.

\cite{segler2018planning} uses the Monte Carlo Tree Search (MCTS) technique, in which molecules are fed into neural networks in the form of extended-connectivity fingerprints (ECFP4s) during the training phase, and two separate but similar neural networks are trained for reaction recommendation and rollout process, where the rollout process comprises a smaller set of rules to ensure a quick assessment of the synthetic accessibility of a compound. The result of the rollout network is then back-propagated to update the positional values of compounds along the pathway. The expansion policy network then chooses the most promising node (compound) to find its $k$-best transformations and sorts the reactions in those transformations using an in-scope filter that gives out probability scores for each reaction.

It is noteworthy that benchmarking retrosynthesis systems can be difficult and subjective, as most studies have been based on a small, publicly available reaction database~\cite{lowe2012extraction}, while some more comprehensive collections of data are only commercially available, and chemists have varying ideas about the judgment of synthesis routes. NLP models use an exact match with the existing literature to define correctness, which is favored for large-scale autonomous benchmarking. However, in real-life scenarios, organic chemists can accept multiple chemically equivalent precursors, as long as the synthons they are derived from are the same. This pursuit of mere memorization of the exact same answer may not only hinder the flexibility of the system but may also undermine the ability of retrosynthesis systems to deliver better results. For practical purposes, double-blind A/B tests on complete routes by experts and laboratory experiments are more convincing, yet anecdotal and not scalable.

Beyond the three components in a retrosynthesis system, neural network-based models dedicated to evaluating and clustering of complete retrosynthetic pathways are also proposed to let the system prioritize the most probable synthetic route and differentiate synthetic strategies~\cite{mo2021evaluating}. The scorer function evaluates a route by its learned pathway embedding, which encodes all the product and reaction fingerprints via the long short-term memory (LSTM) model. A proper clustering of pathways into strategies can enable the system to suggest strategically different routes, rather than providing users with trivially modified routes when the strategy itself is undesired.

%TODO main contribution/overview of ChemiRise
\section{Architecture} \label{sec:architecture}

\begin{figure}
	\centering
	\includegraphics[width=0.5\textwidth]{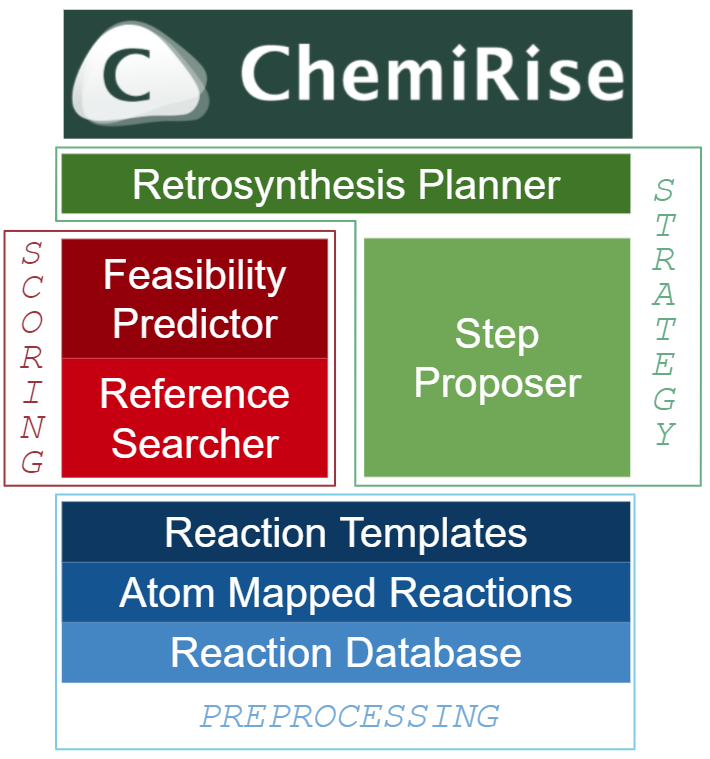}
	\caption{System architecture of ChemiRise, information flow from bottom to the top}
	\label{fig:overview}
\end{figure}

The overall architecture of ChemiRise is shown in \autoref{fig:overview}. ChemiRise is based on data mining and machine learning. The data mining of ChemiRise starts with a reaction database containing millions of experimentally verified reaction records extracted from literatures. The reaction database goes through a series of processing steps which makes the data clean and structured. The most important step is the atom mapping process, which finds the atom correspondences between the reactants and products of each reaction. The atom mapped reactions enable the extracting of reaction templates, which can be seen as a computational description of the underlying chemical process. The structured templates are used to systematically group reactions into clusters and find potential reactions and precursors for any given compound. In summary, the data preprocessing step provides a computational basis for handling chemical reactions.

The retrosynthesis stage, based on the processed reaction database and computational representation of reaction templates, is responsible for generating retrosynthesis plans for given compounds. It consists of two major components: the scoring module and the strategy module.

The scoring module determines the feasibility score of each proposed reaction and is used to filter out chemically unfeasible reactions. To make the results produced by ChemiRise interpretable, the scoring module is \textit{evidence based}: at its heart is an accurate and efficient reaction reference searcher which can quickly find relevant reactions to a proposed reaction.

The strategy module combines the knowledge of the reaction database and the scoring module to generate a route for a given compound. The search state is represented by a directed acyclic graph of explored retrosynthesis space. An adaptive search algorithm is then used to guide the exploration. We trained a graph convolutional neural network to propose the most relevant single retrosynthesis steps for any given compound. The search algorithm repetitively selects an unexplored and commercially unavailable compound in the graph and expands the graph by applying reactions from the single-step reaction proposer. The algorithm succeeds when there exists a plan in the graph in which all starting materials are commercially available.

\section{Data Processing} \label{sec:dataprocessing}

\subsection{Atom Mapping} \label{sec:atommapping}

One major issue in reaction data processing is the lack of atom correspondences in the literature. Although a few reactions cannot be appropriately atom mapped before their mechanisms are elucidated experimentally, most reactions should be correctly atom mapped for the subsequent extraction of reaction templates. More ambiguity is introduced by the fact that only major reactants and products are given in organic reaction databases, and stoichiometry is often ignored. A capable algorithm is then expected to not only balance the chemical equation but also distinguish the role of each reagent, such as pure solvent, a solvent that contributes atoms to the major product, or a catalyst.

Here we employed an empirical scoring function in the two-fold atom mapping process: We first solve the subproblem of atom mapping between a fixed set of reactants and a single product by finding the best scoring mapping through a depth-first search (DFS) algorithm with search space pruning, then finalize the solution using heuristic-based handling of stoichiometry.

\textbf{Scoring function.} The main rationale of the scoring function is to minimize the number of bond modifications. Hence, we assign a score for each kind of matching (atom, bond, etc.) and then sum the score of all matches between the reactants and the product. The scores are empirically determined, approximately following the importance order of (from high to low):

\begin{itemize}
  \item Matched atom element type
  \item Matched atom hybridization
  \item Matched bond count
  \item Bond broken penalty: acyl chloride, disulfide, acid anhydride, ester, etc.
\end{itemize}

\textbf{Pruning.} Even with a fully correct scoring function, the searching for best-scoring atom mapping is still NP-complete since it is no simpler than the graph isomorphism problem. A bare DFS is extremely slow for atom mapping since there are usually many subgraph isomorphisms in reaction graphs. To improve the searching speed, we implemented some comprehensive optimizations and prunings:

\begin{itemize}
  \item \textbf{Partial isomorphism.} %Most reactions only modify a small part of reactants, which means there are full graph isomorphisms between large subgraphs of the reactants and the product. Therefore,  before the atom level searching, we do a preprocess matching to find such partial isomorphisms. We find the local maximums of such partial mappings. Since in this case the subgraph must be fully identical the search space is much lower than the full reaction atom mapping as in that case there may be missing bonds or atoms in the mapping. After such partial isomorphisms are determined, the basic search algorithm is refined: before searching for individual atom mapping assignments, we first try to match large groups of atoms use found partial isomorphisms. Only after all possible partial isomorphisms are tested and the mapping is still incomplete, we fall back to atom level searching for the presumably much smaller set of unmapped atoms.
Most reactions only modify a small part of reactants, which means there are full graph isomorphisms between large subgraphs of the reactants and the product. Therefore, to reduce the workload of the atom-level searching, we firstly perform a partial isomorphism-based search that presumably matches large groups of atoms, which has significantly lower complexity as only fully identical subgraphs are explored, while atom-level searching usually deals with reaction cores that have changes of bonds and atoms.

\begin{figure}
	\centering
	\includegraphics[width=0.9\textwidth]{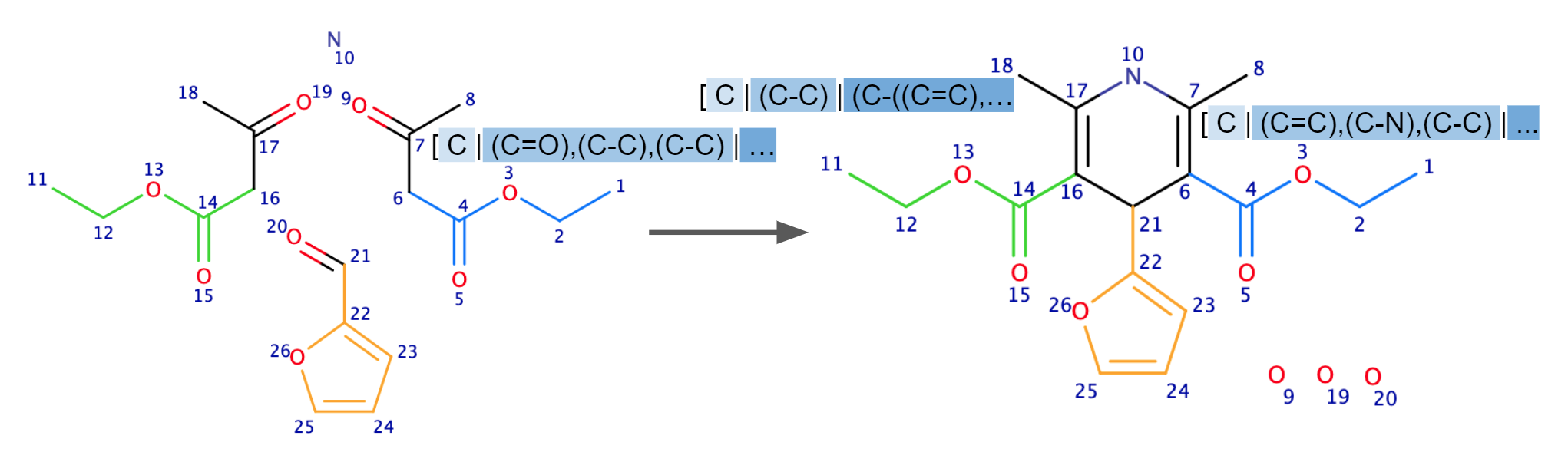}
	\caption{Illustration of the atom mapping process: Subgraphs of the same color are isomorphic, and atoms in unmatched part are layered-hashed. (see \autoref{alg:atomhash})}
	\label{figatommapping}
\end{figure}

  \item \textbf{Layered atom hashing.} During the atom level searching phase, it is possible to prune many improbable possibilities using chemistry knowledge, e.g., atoms can never be matched with atoms of a different element. Furthermore, atoms are more \textit{unlikely} to match atoms with a completely different bonding environment than those with the same environments. This idea is then polished and implemented as the layered atom hashing algorithm (shown in \autoref{alg:atomhash}).

\begin{algorithm}
\SetAlgoLined
\DontPrintSemicolon
\caption{Layered atom hashing}
\label{alg:atomhash}
\KwIn{Compound given as graph $G(V, E)$.}
\KwOut{Layered hash values for each atom which the same hash value corresponds to same atom environment.}
\SetKwFunction{AE}{AtomElement}
\SetKwFunction{BT}{BondType}
\SetKwFunction{Hash}{Hash}
$L \leftarrow \emptyset$\;
\ForEach{$i \in 1...|V|$}{
	$L_{i} \leftarrow \Hash(\AE(V_i))$
}
$H \leftarrow \emptyset$\;
$H_0 \leftarrow L$\;
\ForEach{$i \in 1...\infty$}{
	$L \leftarrow \emptyset$\;
	\ForEach{$j \in 1...|V|$}{
		$E \leftarrow \{ (\BT(E_{j,k}), H_{i-1,k} ) | (j,k) \in E\}$\;
		$L_j \leftarrow \Hash(H_{i-1,j}, E)$\;
	}
   \If{$L=H_{i-1}$}{\Break}
	$H_i \leftarrow L$
}
\Return{$H$}
\BlankLine
\end{algorithm}

The algorithm calculates a layered hash for each of the atoms in the compound. The $i$-th layer hash captures the environment information of the atom within $i$ bonds. The similarity between two layered hashes is defined to be the highest layer where the two hashes are identical. This similarity is then used as the searching priority during the DFS.

\item \textbf{Early stopping.} %During the search, we keep track of the current highest matching score. Then at each depth first search step, we inspect the currently mapped atoms and calculates the potential maximum score if we proceed with current mapping. If this score is less than already recorded high score, we can terminate traversing on current search branch.
During the search, we keep track of the current highest matching score. Then a searching branch can be pruned during a DFS step if the potential maximum score with current atom mapping is less than the recorded high score.
\end{itemize}

\textbf{Stoichiometry.} On top of the above reaction mapping algorithm on fixed reactants, a heuristic-based strategy is used to handle stoichiometry. If the atom mapping algorithm fails to find a mapping using one instance of each reactant, it will try adding one instance of each reactant and apply the addition that leads to the most increase in score. This process is repeated until the mapping process succeeds.
 %We begin with single instance of each reactant. If the reaction mapping algorithm proceeded successfully, we output the result. Otherwise, we try to add a copy of each reactant and check which one leads to the increasing of score. We add the one with the best score improvement and repeat the process until the mapping process succeeded. 

\subsection{Reaction Template} \label{sec:reactiontemplate}

\begin{figure}
	\centering
	\includegraphics[width=0.7\textwidth]{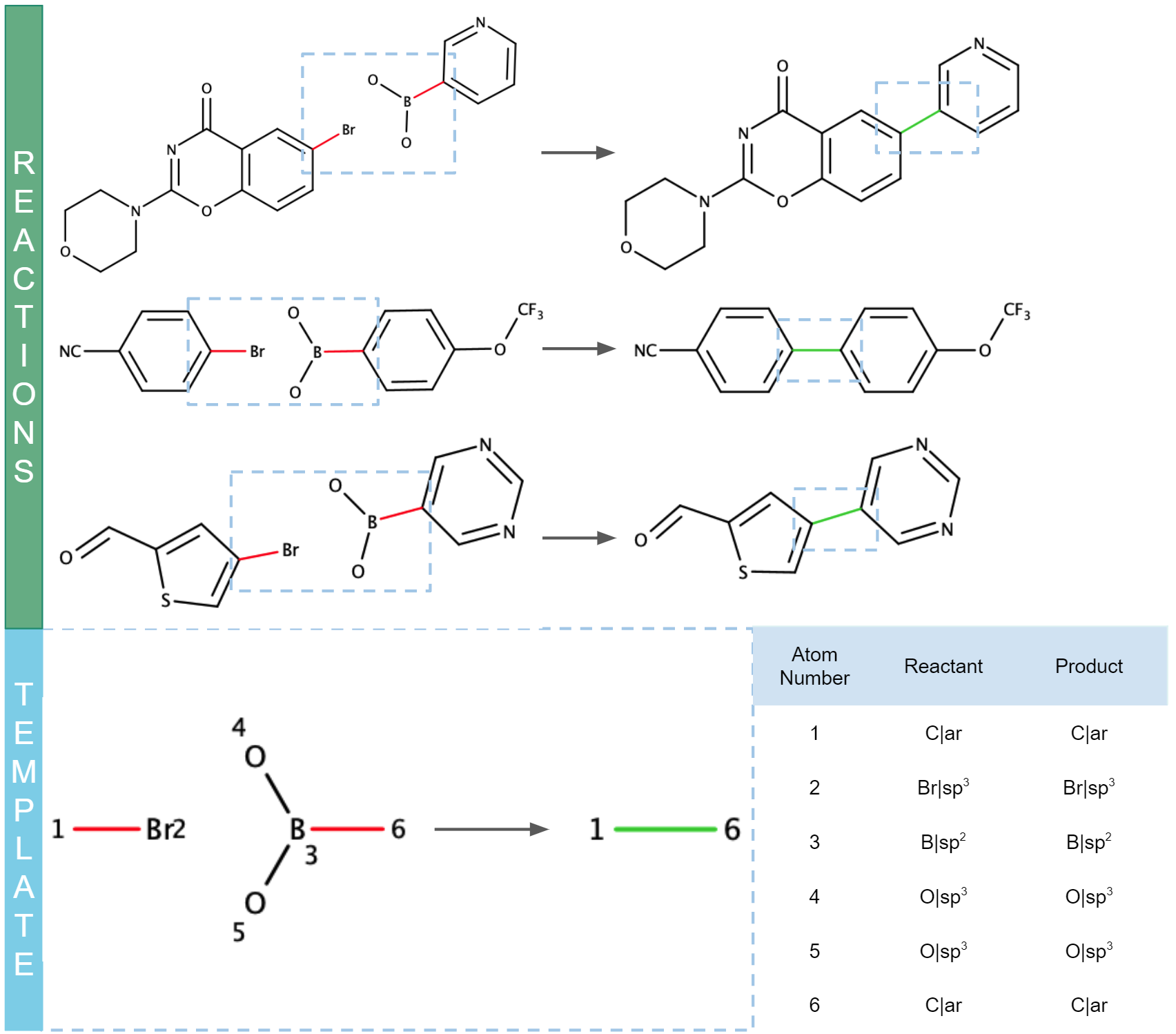}
	\caption{Illustration of the template extraction of the Suzuki coupling reaction}
	\label{figtemplate}
\end{figure}

After the atom correspondences of reactions are determined, \textit{reaction templates} can be extracted from these atom mappings. An illustration of the extraction process for the Suzuki coupling reaction is shown in \autoref{figtemplate}. Atoms without changes in bonding or hybridization states are removed from the mapped reaction to yield a reaction graph that contains minimal information of the graph-level changes in the reaction. Each atom in a template is labeled with its atom type, which consists of information about its element, hybridization type, and connected bonds. Atom correspondences are also kept in the reaction template. A reaction template can be seen as a compact representation of a corresponding chemical reaction. They enable efficient computational procedures to find potential reaction sites on a product and be applied upon those sites to find retrosynthesis precursors of the step using the reaction it represents.

Reactions of the same reaction templates are then grouped to form \textit{reaction clusters}. These clusters will then be used in subsequent sections for collecting statistical information for each reaction type.

\section{Retrosynthesis Planning} \label{sec:retrosynthesis}

\subsection{Reaction Feasibility Predictor} \label{sec:feasibilitypredictor}

\begin{figure}
	\centering
	\begin{subfigure}{0.45\textwidth}
		\centering
		\includegraphics[width=0.5\textwidth]{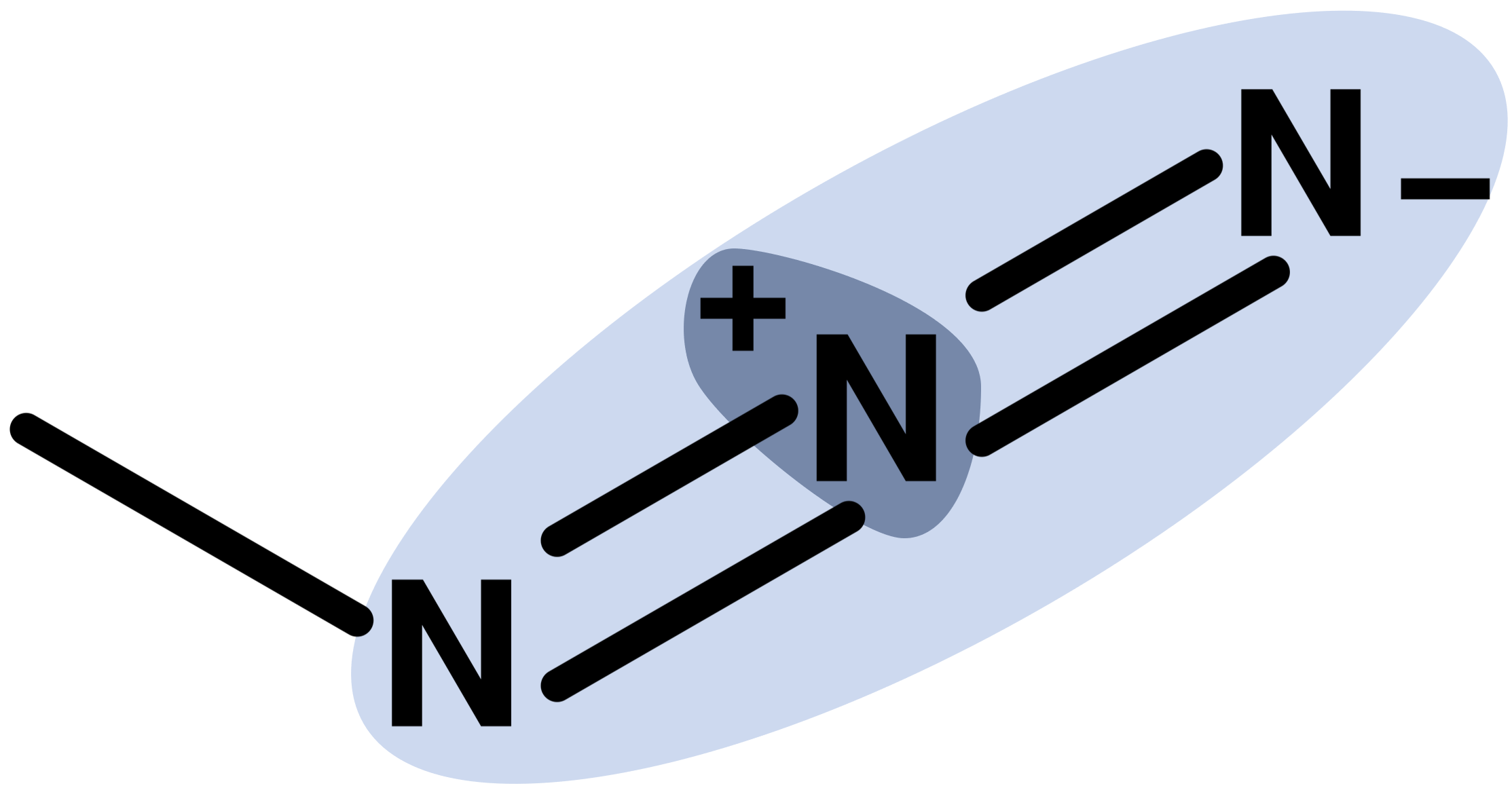}
		\caption{Atom-centric functional group extraction of azides, centered at the middle nitrogen atom.}
	\end{subfigure}
	\begin{subfigure}{0.5\textwidth}
		\centering
		\includegraphics[width=0.5\textwidth]{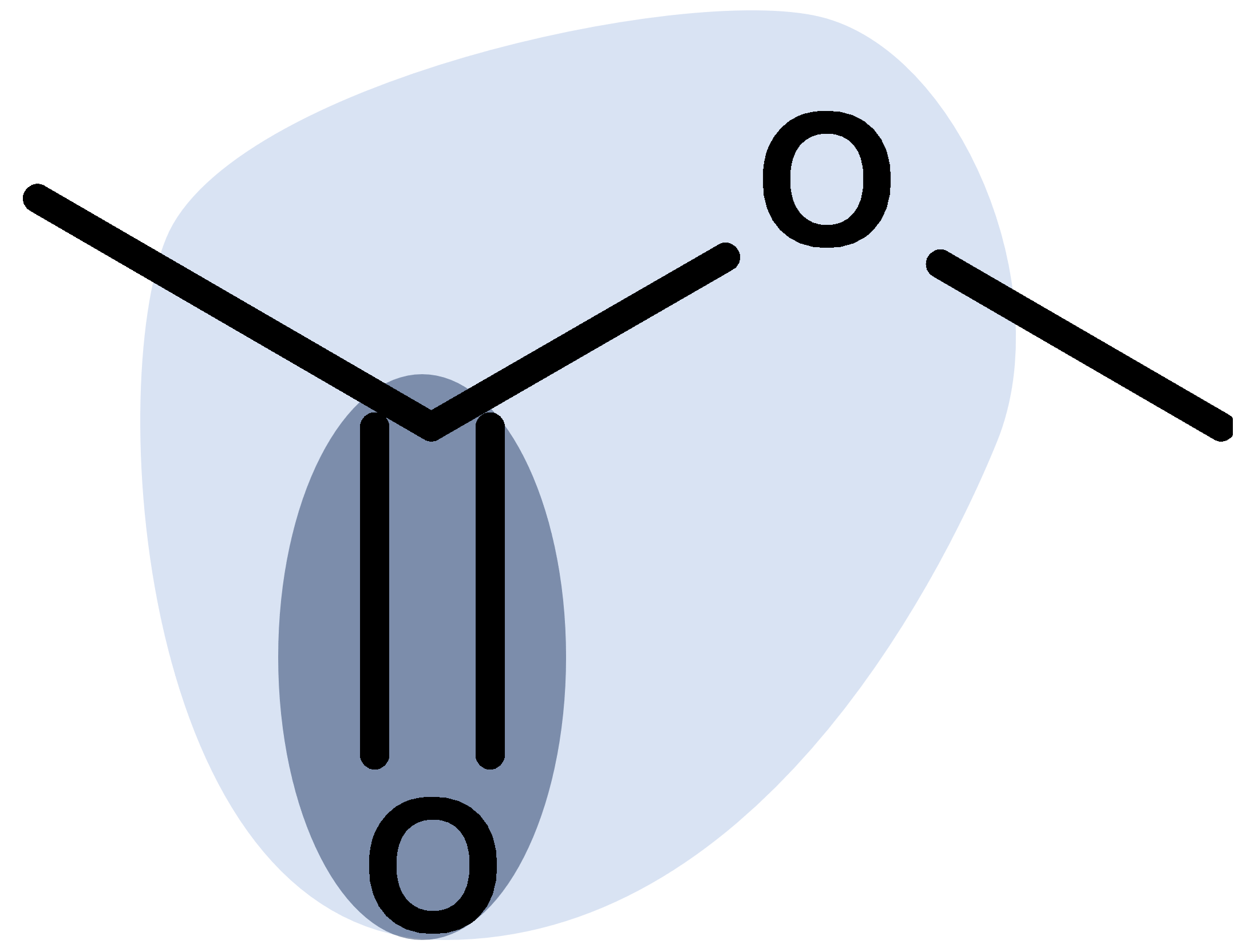}
		\caption{Bond-centric functional group extraction of esters, centered at the double bond in the carbonyl group, with the trivial $sp^3$ carbon neglected.}
	\end{subfigure}
	\hfill
	\caption{Automatic \textit{atom group} extraction}
	\label{figfuncgroup}
\end{figure}

\textbf{Functional groups.} To determine the feasibility of a chemical reaction, functional groups in the reactants need to be evaluated by their compatibilities with the desired reaction condition. Instead of manually maintaining a list of such functional groups used in previous studies~\cite{klucznik2018efficient}, we utilized automatically extracted functional group templates called \textit{atom groups}. Two types of atom groups are defined: an \textit{atom centric} group is defined as a central atom with its directly bonded neighbor atoms, which are extracted for all heavy atoms except carbon; a \textit{bond centric} group is defined as a central bond consisting of both atoms at its ends, with their directly bonded neighbor atoms. \textit{Bond centric} groups are extracted for all non-single bonds. Examples of such atom groups are shown in \autoref{figfuncgroup}. Atoms are assigned with attributes such as aromatic, bonded to non-$sp^3$ atoms, located in a ring, etc., to assist the discrimination of the stability and compatibility of atom groups, while some trivial $sp^3$ carbons are removed from the atoms in the group in lieu of having chemically redundant atom groups.

\textbf{Feasibility score.} We calculate the feasibility score of a reaction $R$ by two parts - the reactivity and compatibility. For determining reactivity, we use an empirically-determined confidence score based on the size and frequency of the corresponding reaction template. For functional group compatibility, we extract the functional groups in the reactants. Assuming independence of each functional group $g_i$, a Naive Bayes-like estimator is used to determine the compatibility score:

\begin{equation}
  P(R|\textbf{g}) = \frac{P(\textbf{g}|R)P(R)}{P(\textbf{g})} = \prod{\frac{P(g_i | R)P(R)}{P(g_i)}}
\end{equation}
where the probabilities $P(g_i|R)$, $P(R)$ and $P(g_i)$ are estimated by the occurrence frequencies of the corresponding entities using statistics in the reaction template cluster and the reaction database.

\textbf{Selectivity.} To determine the selectivity of a proposed reaction, we first find all potential forward reaction sites $\{s_i\}$ by applying reaction templates to the reactants, and calculcate the feasibility score of each reaction site, denoted as $F(s_i)$. The final feasibility score is simply defined to be the fraction of the score of the main site (denoted as $s_0$):

\begin{equation}
  F(s_0|\{s_i\}) = \frac{F(s_0)}{\sum_i{F(s_i)}}
\end{equation}

\textbf{Reference Searcher.} Literature references are crucial information for chemists to judge the viability of a proposed reaction that can resolve common concerns such as the fragility of certain functional groups or the reactivity of the compound by showing the existence of similar features in published reactions. The task is thus to find the most relevant reactions in the database for given queries, i.e. reaction steps generated by the retrosynthesis planner. We use specifically designed features for the reaction finding process, which include:

\begin{enumerate}
  \item Reaction template ID, including extended reaction template ID
  \item Reaction site features, including site features for other regioselectivity-competing sites
  \item Environmental atom group features
  \item CAS similarity features (CAS structure descriptors used in similarity calculations)~\cite{dittmar1983cas}. 
\end{enumerate}

To find the most similar reactions to a proposed reaction, the system calculates the distance of its feature $f$ to the feature $g$ of each reaction in the reaction database:

\begin{equation}
  dist(f, g) = \theta_i^{sym}(\sum_i f^{sym}_i - g^{sym}_i)^2 + \theta_i^{asym}(\sum_i \max (0, f^{asym}_i - g^{asym}_i))^2
\end{equation}
where $\theta$ is predefined weighting for each considered feature.

The features are categorized into symmetric ($f^{sym}, g^{sym}$) and asymmetric ($f^{asym}, g^{asym}$) parts. The rationale behind such categorization is to assess the compatibility of reactions properly, as chemists often raise questions about the compatibility of an atom group in the proposed reaction scheme if such an atom group never appeared before in similar reactions. As a result, in terms of the environmental atom group features, it is acceptable for the reference reaction to have excessing features (atom groups), but unacceptable for them to have missing features (atom groups). 

An inverted index-based database system is implemented to speed up these reference searches, as simple iteration is unfeasibly slow due to the size of the reaction database.

\subsection{Reaction Proposer} \label{sec:reactionproposer}

The reaction proposer is the basis of the retrosynthesis planning algorithm. For a given compound, instead of directly outputting candidate reactants, the proposer uses a graph convolution-based method to search for the best reaction templates in a retrosynthesis step and applies the templates if their reaction site is found to produce full chemical reactions to be proposed. This implementation strictly ensures reaction sites are compatible with templates and enables the use of the reaction feasibility predictor by having discrete reaction templates.

The reaction proposer is based on a graph convolutional neural network similar to the one in Chemi-Net, our previously published neural network architecture~\cite{liu2019chemi}. The input compound is first quantized as a molecular graph $G(V, E)$, with vertices $V$ corresponding to the atoms and edges $E$ corresponding to the bonds. The initial atom embedding is defined through an embedding lookup table of the atom element. Each of the graph convolution layers consists of an aggregation of embeddings of atoms bonded to each atom:

\begin{equation}
  A_i^k = f(Reduce(\{W^k(Concat\{A_i^{k-1},A_j^{k-1},P_{i,j}\})|(i,j)\in E\}))
\end{equation}
where $W^k$ is the learned weight matrix for layer $k$, $P_{i,j}$ is the bond embedding between atom $i$ and $j$, $f(\bullet)$ is the non-linear activation function, $Reduce(\bullet)$ is the set reduction function which reduces a variable number of embeddings to a fixed size embedding. We use the set2set~\cite{vinyals2015order} network for the set reduction.

The final embedding of the entire compound is defined as the set reduction of all the atom embeddings:

\begin{equation}
  E = Reduce(\{A_i^K| i \in V\})
\end{equation}
where $K$ is the number of graph convolutional layers.

The outputs of the neural network are pointwise scores of each reaction template. The ranking of such scores is the output of the reaction proposer. As in \autoref{sec:reactiontemplate} we assign a unique reaction template to each reaction in the reaction database.

The simplest method to realize such scores is to treat the problem as a multi-class classification and use the standard cross-entropy loss function for training. However, treating the reaction templates merely as discrete classes loses the molecular structure data of the reaction templates. For example, the applicability of a Suzuki coupling using bromide should be very similar to the one using iodine, albeit they are of different reaction templates. Hence, we use \textit{template embedding} to enable the model to learn the knowledge of the similarity between reaction templates. Since the reaction template is composed of multiple compounds, i.e., reactants and products, to perform the aforementioned molecular graph quantization, pseudo edges are added between every pair of mapped atoms in the reactant and product, which makes the template a connected graph. The template embedding is then similarly defined as the graph convolutional network embedding of the resulting template graph.

After the template embedding is calculated, the final proposer score for each reaction template is defined as the dot product between the embedding of the target compound $E$ and the template embedding of each reaction template $T_i$:

\begin{equation}
  P_i = E \cdot T_i
\end{equation}

\subsection{Retrosynthesis Planner} \label{sec:retrosynthesisplanner}

\begin{figure}
	\centering
	\begin{subfigure}{0.95\textwidth}
		\centering
		\includegraphics[width=\textwidth]{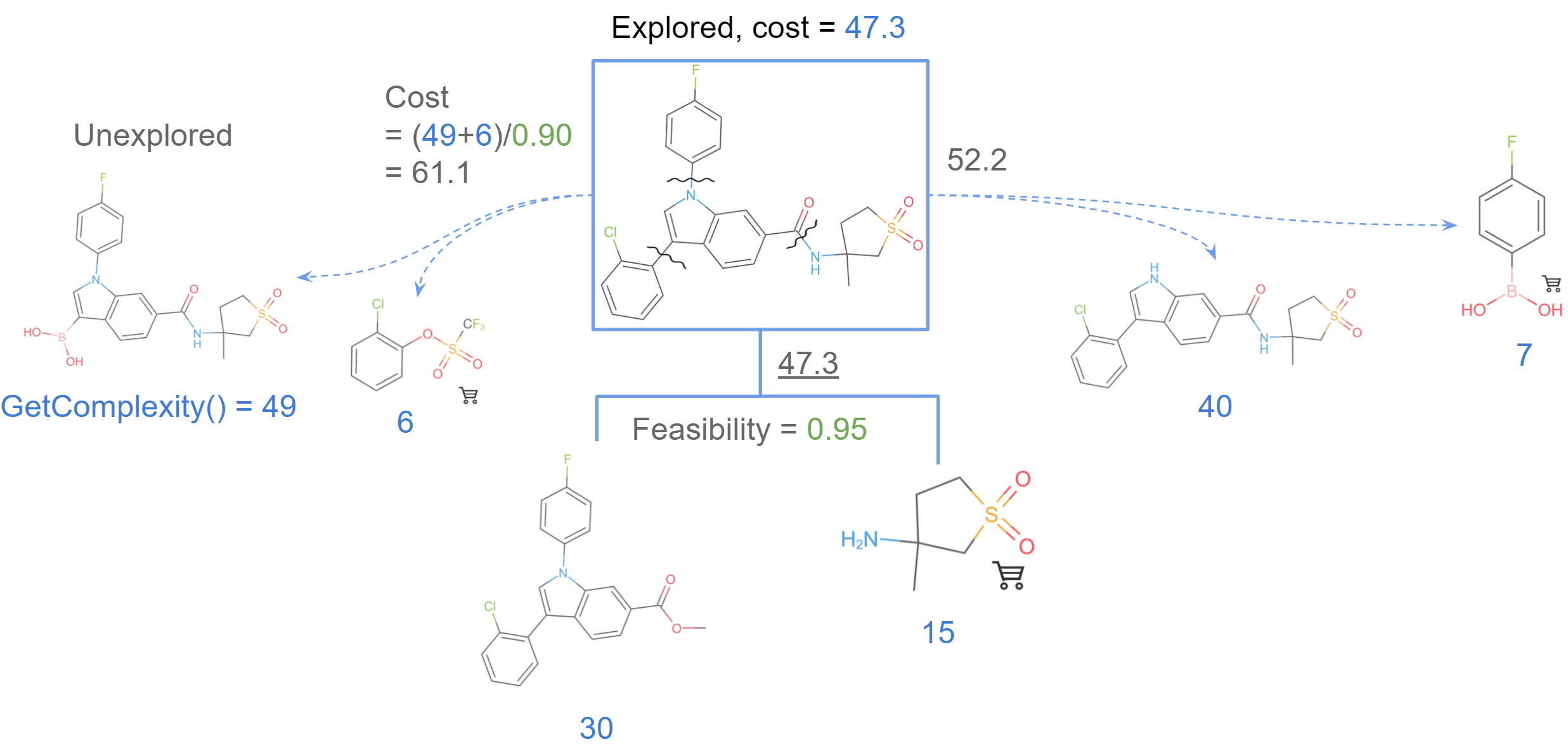}
		\caption{For unexplored compounds, an estimator function (GetComplexity(\textit{compound})) is called to have an educated guess. For explored compounds with multiple retrosynthesis routes, the lowest cost is updated as the cost of the compound.}
	\end{subfigure}
	\begin{subfigure}{0.95\textwidth}
		\centering
		\includegraphics[width=\textwidth]{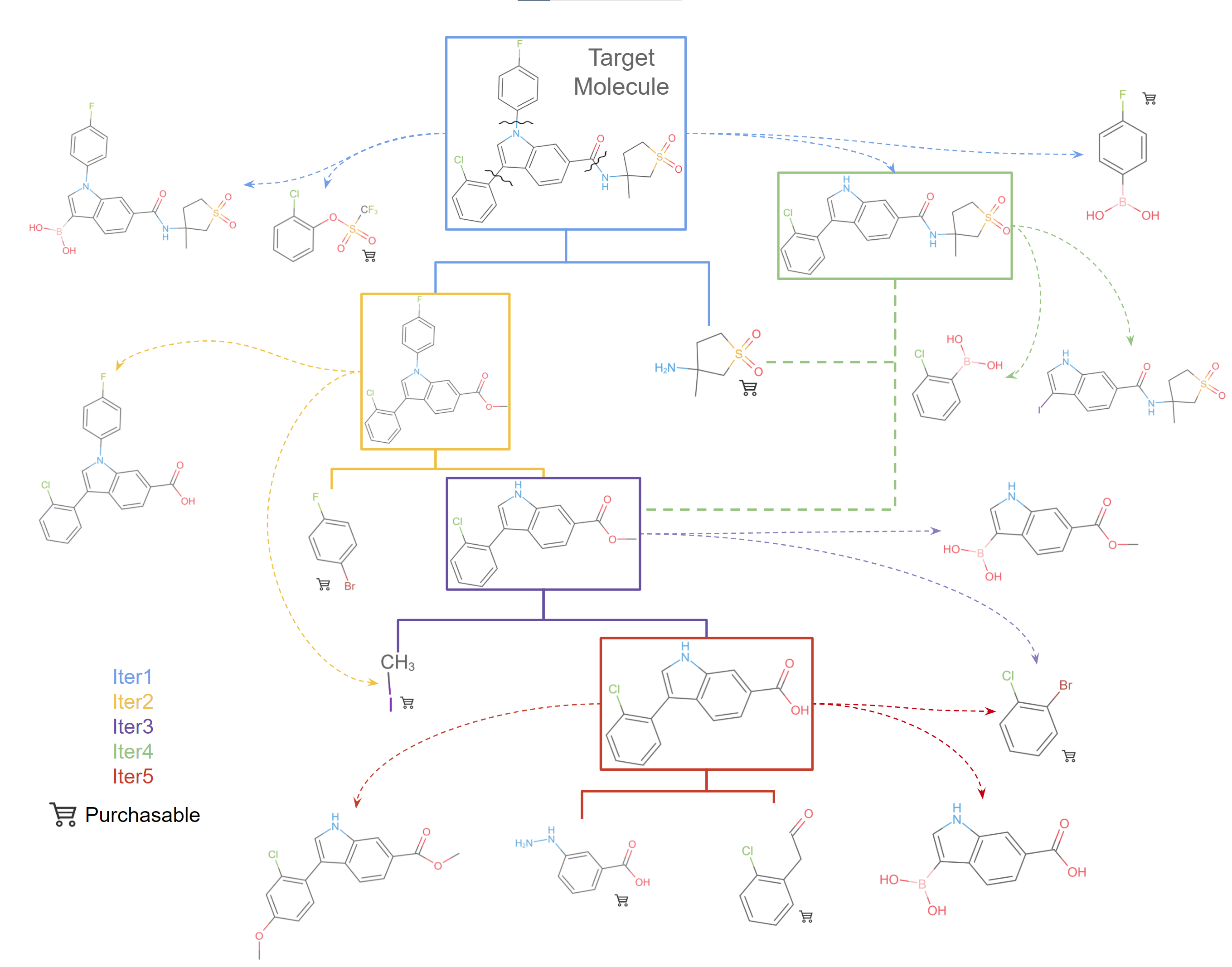}
		\caption{A complete search of retrosynthesis routes. Five iterations are marked with different colors, and the target compound to explore in each iteration is in a square of corresponding color. Rectangle lines indicate the best reaction found in each iteration, and solid lines indicate the final route.}
	\end{subfigure}
	\caption{An illustrative process showing five expansion iterations of the ChemiRise retrosynthesis engine.}
	\label{figexpansion}
\end{figure}

The retrosynthesis planning process uses a directed acyclic graph (DAG)-based searching algorithm to efficiently search the synthesis space. The searching procedure is illustrated in \autoref{figexpansion}. \autoref{figexpansion}a shows the exploration of a target compound in a single iteration, in which compatible templates are applied to the molecule to yield its possible precursors. The cost of each reaction is estimated by the compound complexity estimator and reaction feasibility score(\autoref{alg:cost}). The step with the lowest cost is then chosen (rectangular lines), and the updated cost is fed back to other compounds along the synthesis path in the graph. \autoref{figexpansion}b presents a complete route searching process with five iterations. Each iteration expands the compound in the frame of its corresponding color. During the search, the updated costs may alter the preference of expansion location after each iteration, as iteration four (Iter4) demonstrates, so that the algorithm may choose compounds in other routes that are not the most deeply explored. It helps the searcher to circumvent traps where an estimated low-cost compound turns out to be expensive to synthesize after several expansions within the route. In \autoref{figexpansion}b, after iteration four the cost of the new route is updated to be higher than the original one, so iteration five (Iter5) is executed back on the original route, which completes the route by finding two purchasable precursor compounds.

A rigorous description of the main retrosynthesis planning algorithm is shown in \autoref{alg:planning}. The best reactant to be explored in each iteration is defined to be the reactant whose expansion may lead to the maximum reduction of the cost of the target compound. In practice, an estimator function is used to assign heuristic values on compounds, which can be roughly interpreted as the synthetic cost or complexity of the molecule.

\begin{algorithm}
\SetAlgoLined
\DontPrintSemicolon
\caption{Retrosynthesis planning algorithm}
\label{alg:planning}
\KwIn{Target compound $m$.}
\KwOut{The best retrosynthesis route $R$.}
\SetKwFunction{Route}{GetFeasibility}
\SetKwFunction{Propose}{ProposeReactions}
\SetKwFunction{Cost}{Cost}
\SetKwFunction{SAScore}{GetComplexity}
\tcp{Compound set}
$S = \{ m \}$\;
\ForEach{$iter \in 1...iter\_limit$}{
	\tcp{Find current best route}
	$r \leftarrow \argmin_r { \Cost(r) }$\;
	\tcp{Find most complex compound in the route}
	$c \leftarrow \argmax_{c\in r} { \SAScore(c) }$\;
	$R_c \leftarrow \Propose(c)$\;
	\ForEach{$r \in R_c$}{
		\ForEach{$reactant \in r$}{
			$S \leftarrow S \cup \{ reactant \}$\;
		}
	}
}
\uIf{$\exists route \in S$}{
	\Return{$route$}
}
\Else{
	\Return{$\emptyset$}
}
\BlankLine
\end{algorithm}

The cost of a compound is calculated by its complexity if no retrosynthesis step has been explored from it yet, otherwise, the cost is determined by the cost of the reactants and the feasibility of the reaction. (See \autoref{alg:cost})

\begin{algorithm}
\DontPrintSemicolon
\caption{The cost function}
\label{alg:cost}
\KwIn{Compound $m$.}
\KwOut{The cost to synthesize compound $m$.}
\SetKwFunction{LScore}{GetFeasibility}
\SetKwFunction{SAScore}{GetComplexity}
\uIf{$R_m = \emptyset$}{
	\Return{$\SAScore(m)$}
}
\Else{
	\Return{$\min_{r\in R_M}{  \sum_{reactant \in r}{Cost(r)} / \LScore(r) }$}
}
\end{algorithm}

To improve performance, the retrosynthesis engine is fully distributed over all available CPU cores or across machines. We maximize the concurrency by allowing multiple iterations to be simultaneously running.

\section{Experimental Results} \label{sec:results}

\subsection{Atom Mapping}

We compared our results with industry-standard packages: Marvin and Indigo. We used the following two evaluation metrics to test the accuracy of our atom mapping algorithm:

\begin{itemize}
  \item Complete mapping rate, which evaluates whether all product atoms are mapped in the atom mapping output. This measures recall of the mapping process, as missing atom mappings come from the failure of the algorithm.
  \item Average number of C-C bonds broken in the output atom mapping, which measures precision as it is unlikely for a ground truth reaction to break much more C-C bonds. A large number of such breakages usually means wrong atom mappings.
\end{itemize}

The results are shown in \autoref{table:atommapresult}. Our algorithm performed noticeably better in terms of precision and recall.

\begin{table}
\centering
\caption{Evaluation results of atom mapping algorithms}
\begin{tabular}{ccc} \toprule
  Algorithm & Complete mapping rate & Average number of C-C bond broken \\ \midrule
  ChemiRise & 73\%                  & 0.28 \\ 
  Marvin    & 63\%                  & 0.30 \\ 
  Indigo    & 57\%                  & 0.37 \\ \bottomrule
\end{tabular}

\label{table:atommapresult}
\end{table}

\subsection{Reaction Proposer}

We compare the accuracy of our reaction proposer with previous work~\cite{coley2017computer} using the dataset in~\cite{lowe2012extraction}, which comprises about 50k reactions. For each reaction in the benchmark dataset, we use our reaction proposer to predict potential reactions and precursors of the product. We then check if the original reaction occurred in the top-K ranked predictions. This experiment tests the ability of the model to imitate human proposed reactions. The results are shown in \autoref{table:recommenderresult}. It is apparent that our graph convolutional neural network model has better prediction accuracy.

\begin{table}
\centering
\caption{Evaluation results of reaction proposers}
\begin{tabular}{ccccccc} \toprule
  \multirow{2}{*}{Model} & \multicolumn{6}{c}{Top-N Accuracy (\%)} \\ \cmidrule{2-7}
               & 1      & 3      & 5     & 10    & 20    & 50    \\ \midrule
  ChemiRise    & 43.8   & 62.1   & 70.1  & 78.3  & 85.0  & 90.5  \\ \midrule
  Coley et al. & 37.3   & 54.7   & 63.3  & 74.1  & 82.0  & 85.3  \\ \bottomrule
\end{tabular}
\label{table:recommenderresult}
\end{table}

\subsection{Retrosynthesis}

The evaluation of a computational retrosynthesis is always challenging since it is impossible to produce a "ground truth" route nor to judge whether a generated route is the best route. This is because many routes are possible for a given target and different routes may suit different situations of the synthesis. For example, a process chemist who is seeking an optimal procedure for a given target can accept uncertainly of the feasibility of the proposed reactions for lower costs in mass production, while a medicinal chemist usually can tolerate longer and uneconomical routes but want to minimize the uncertainty of the reactions.

Hence, we resort to expert scoring of the proposed routes. The score is formulated as several levels:

\begin{enumerate}[A:]
  \item The route can be implemented in the laboratory directly or with only minor adjustments. Minor changes include change of protecting groups 
  \item The route has a reasonable strategy but requires major changes to be implemented in the laboratory.
  \item The route is unreasonable and provides little value for the chemist.
  \item The algorithm does not output a complete route to purchasable compounds at all.
\end{enumerate}

Synthesis routes of compounds in \autoref{table:retrosynthesisresult} are listed in Supporting Information.
\begin{table}
\centering
\caption{Evaluation results of retrosynthesis case study}
\begin{tabular}{c|cccc} \toprule
  Target compound & Result steps & Time & Iterations & Expert Score \\ \midrule
  \includegraphics[width=4cm,trim=0 -5 0 -5]{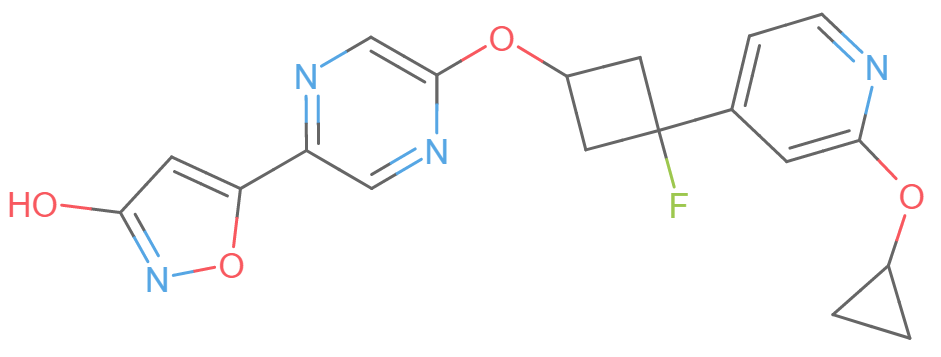} & 6  & 30s & 500 & A \\ \midrule
  \includegraphics[width=4cm,trim=0 -5 0 -5]{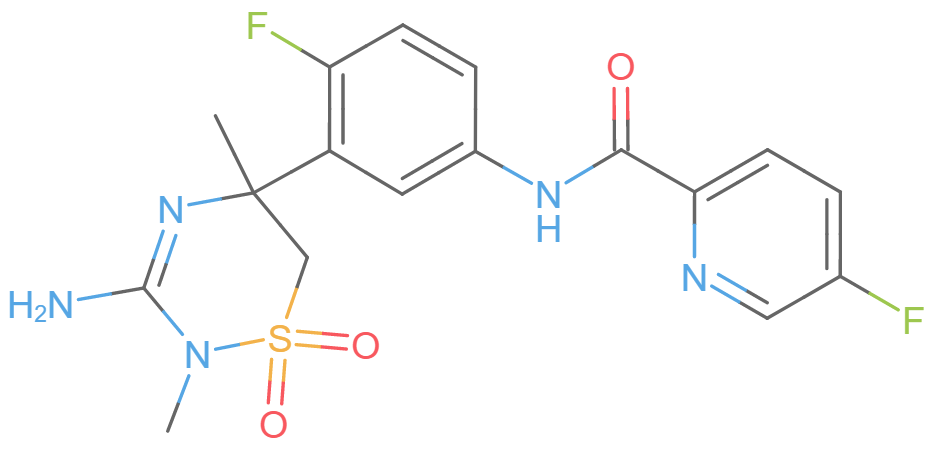} & 6 & 50s & 500 & A \\ \midrule
  \includegraphics[width=4cm,trim=0 -5 0 -5]{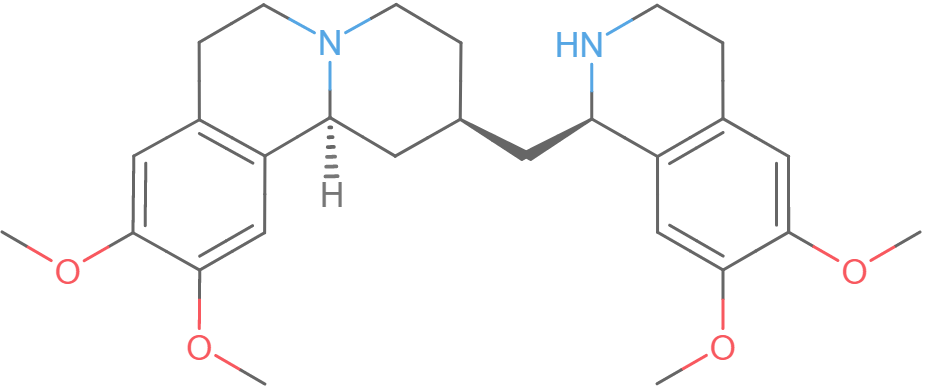} & 5  & 12s & 89  & B \\ \midrule
  \includegraphics[width=2cm,trim=0 -5 0 -5]{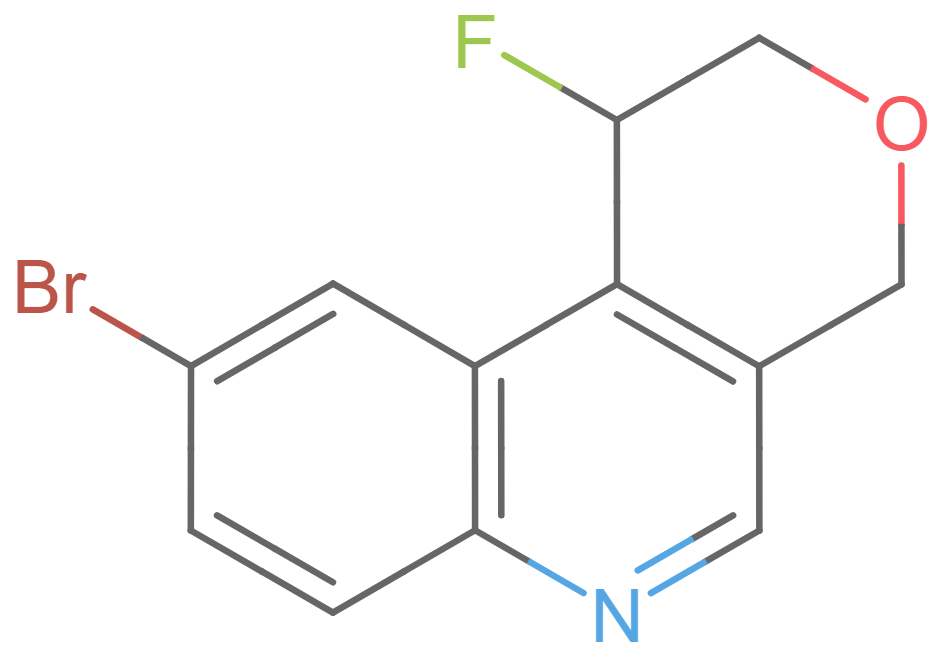}  & 6  & 10s & 49  & C \\ \midrule
  \includegraphics[width=4cm,trim=0 -5 0 -5]{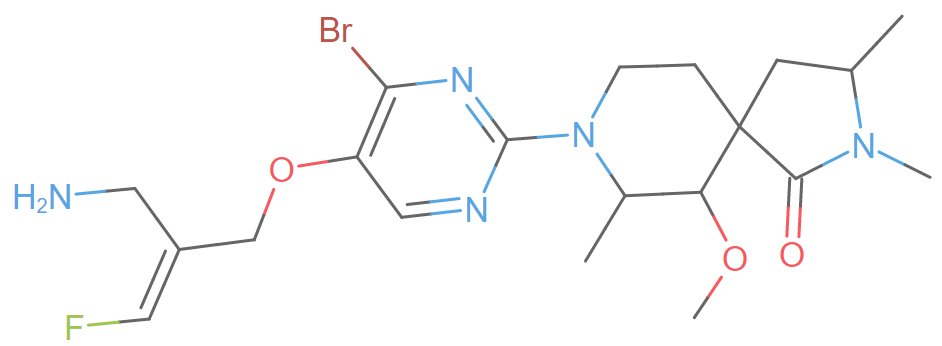} & 10 & 21s & 163  & A \\ \midrule
  \includegraphics[width=2cm,trim=0 -5 0 -5]{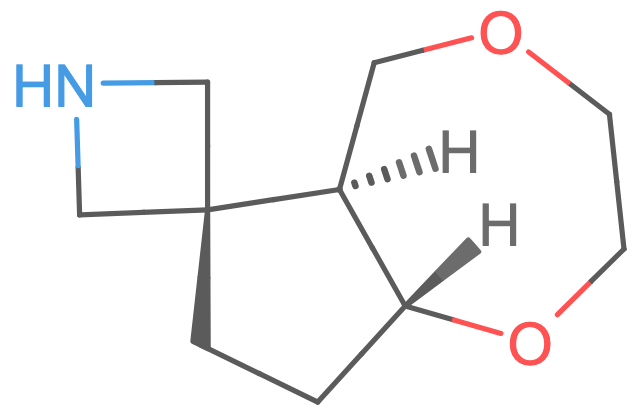} & 9  & 15s & 293 & C \\ \midrule %rerun result not the same as Ji's
  \includegraphics[width=4cm,trim=0 -5 0 -5]{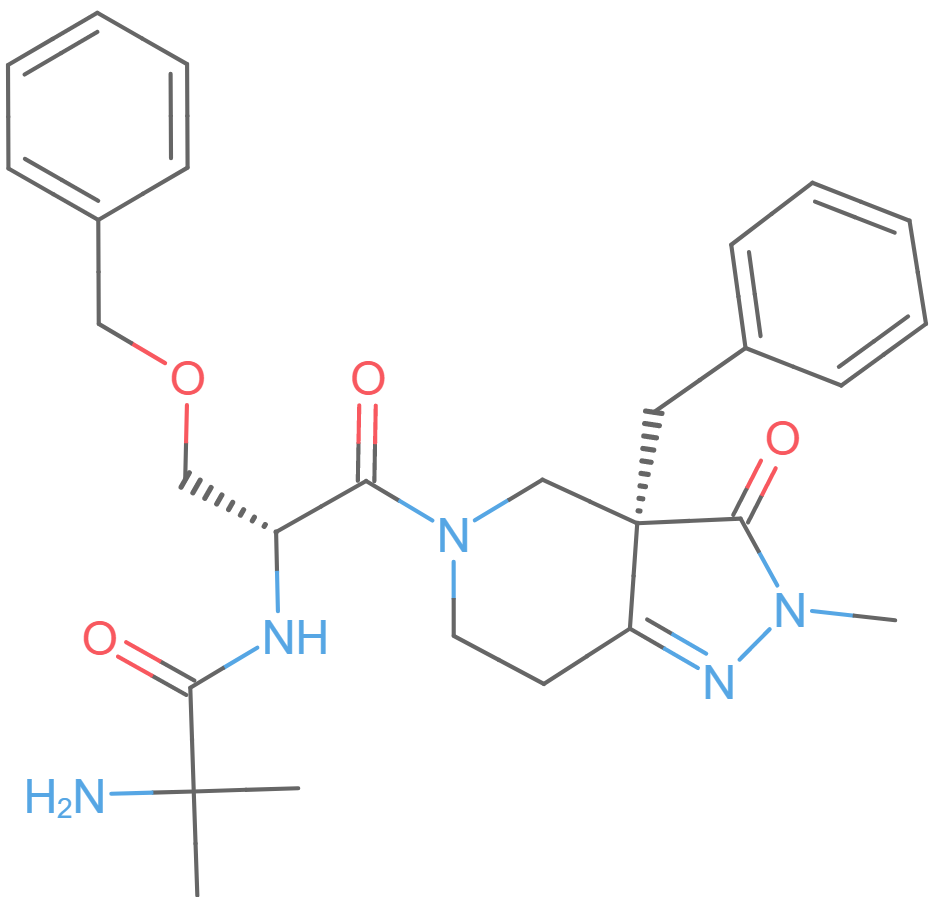} & 4  & 10s & 74  & A \\ \bottomrule
 \end{tabular}
\label{table:retrosynthesisresult}
\end{table}

\begin{figure}
	\centering
	\begin{subfigure}{0.9\textwidth}
		\centering
		\includegraphics[width=\textwidth]{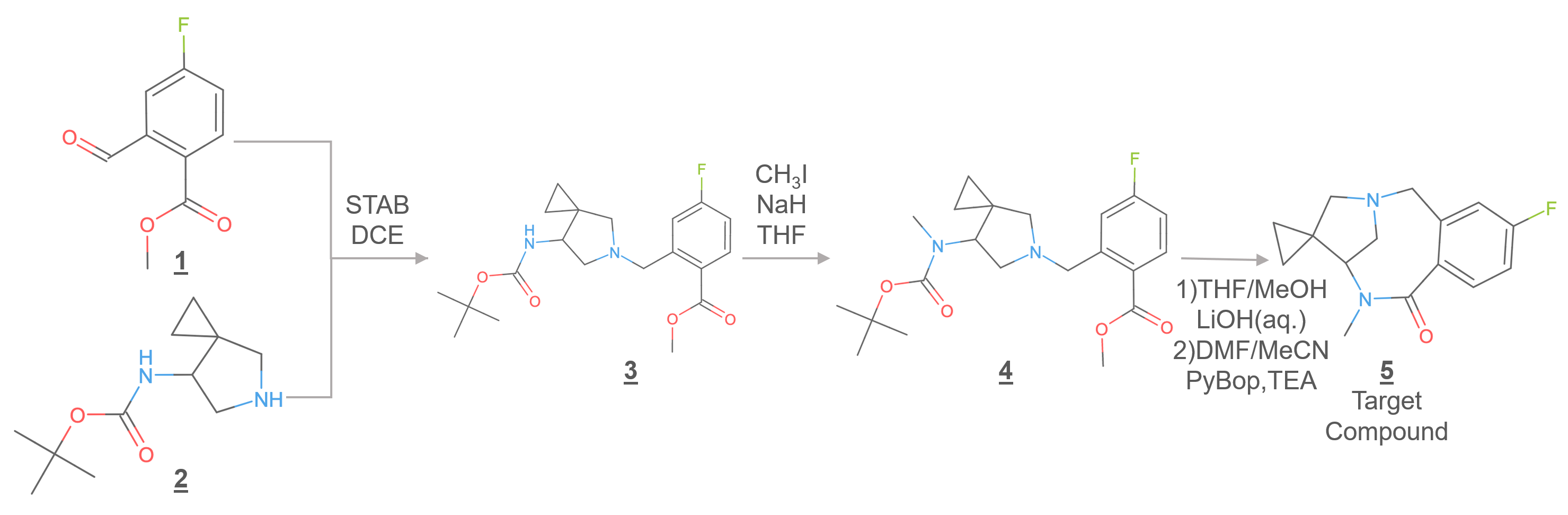}
		\caption{}
	\end{subfigure}
	\begin{subfigure}{0.95\textwidth}
		\centering
		\includegraphics[width=\textwidth]{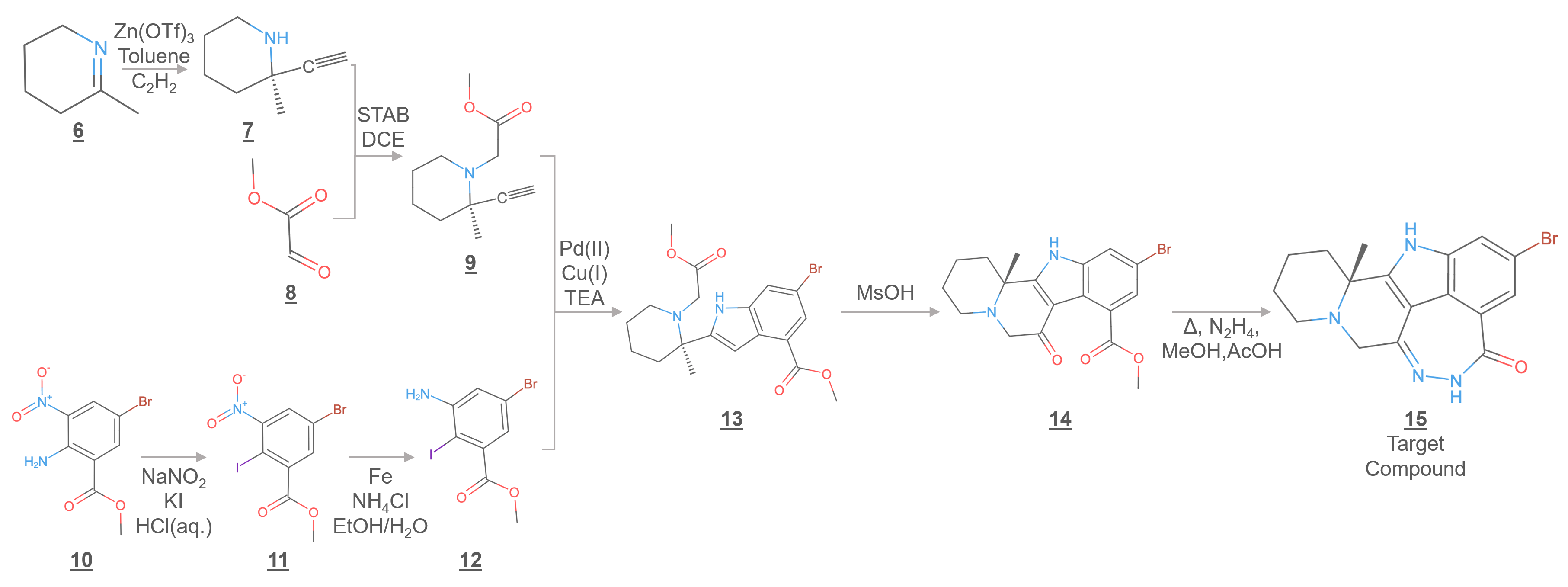}
		\caption{}
	\end{subfigure}
	\begin{subfigure}{0.8\textwidth}
		\centering
		\includegraphics[width=\textwidth]{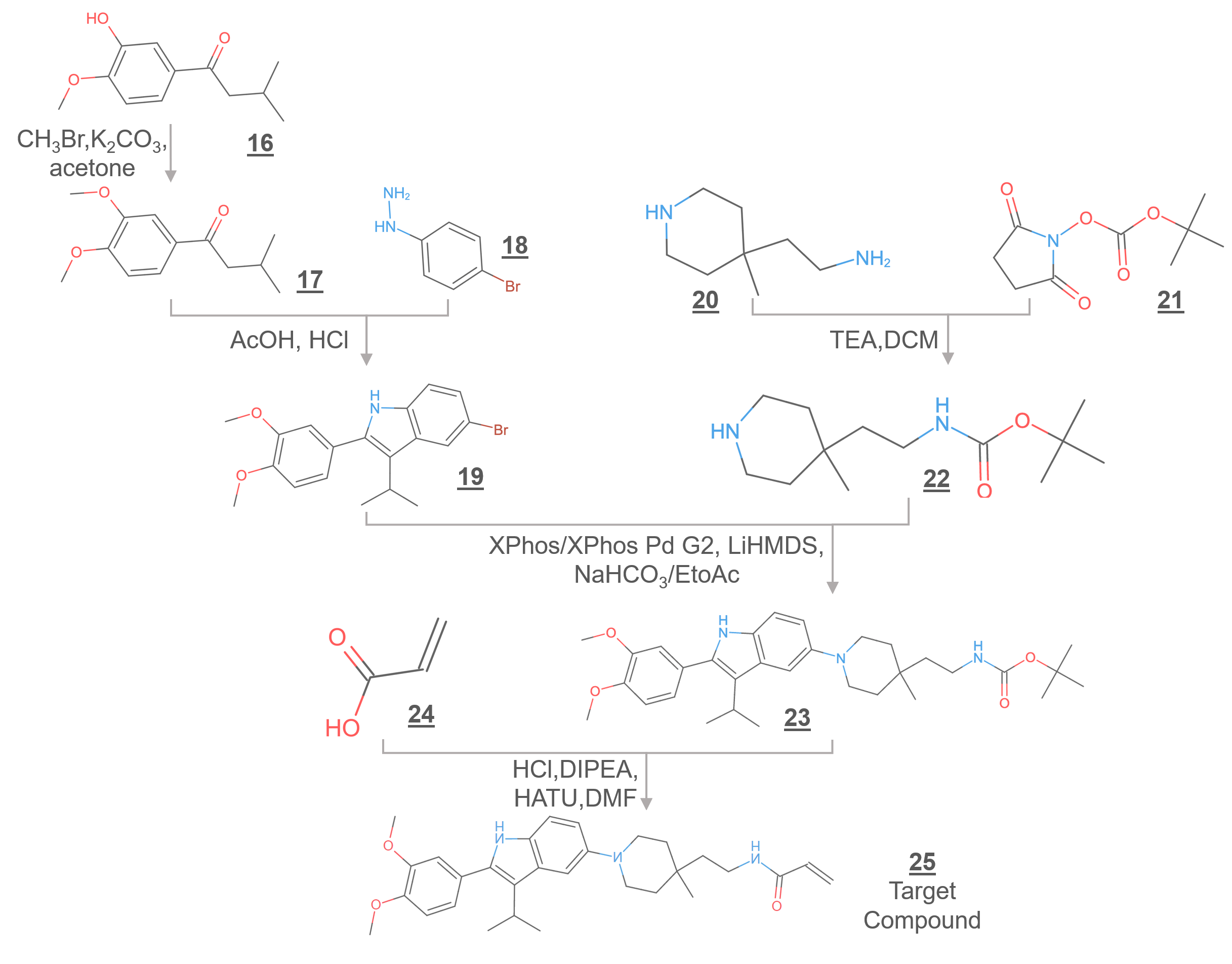}
		\caption{}
	\end{subfigure}
	\caption{Synthesis routes generated by ChemiRise, reaction conditions and reagents are taken from references.} 
	\label{figroutes}
\end{figure}

Furthermore, the expert demonstrated several routes that exemplify the reliability of the system (\autoref{figroutes}). All routes shown are generated by ChemiRise in a single run without additional expert input such as manually choosing reactions. The ability to search in a much larger database of purchasable chemicals than human experts also reduces the repetitive effort to synthesize some building blocks. For instance, the purchasability of the spiro compound 2 leads to a simple three-step synthesis of compound 5. Also, knowing available compounds with convenient, already-in-place substituents, the system chooses different strategies for the synthesis of the indole ring in \autoref{figroutes}b and \autoref{figroutes}c, with correct modifications performed without selectivity or fragility issues, such as the iodination/reduction process from compound 10 to 12, and the protection/deprotection from compound 20 to 25. 

\begin{figure}
	\centering
	\includegraphics[width=\textwidth]{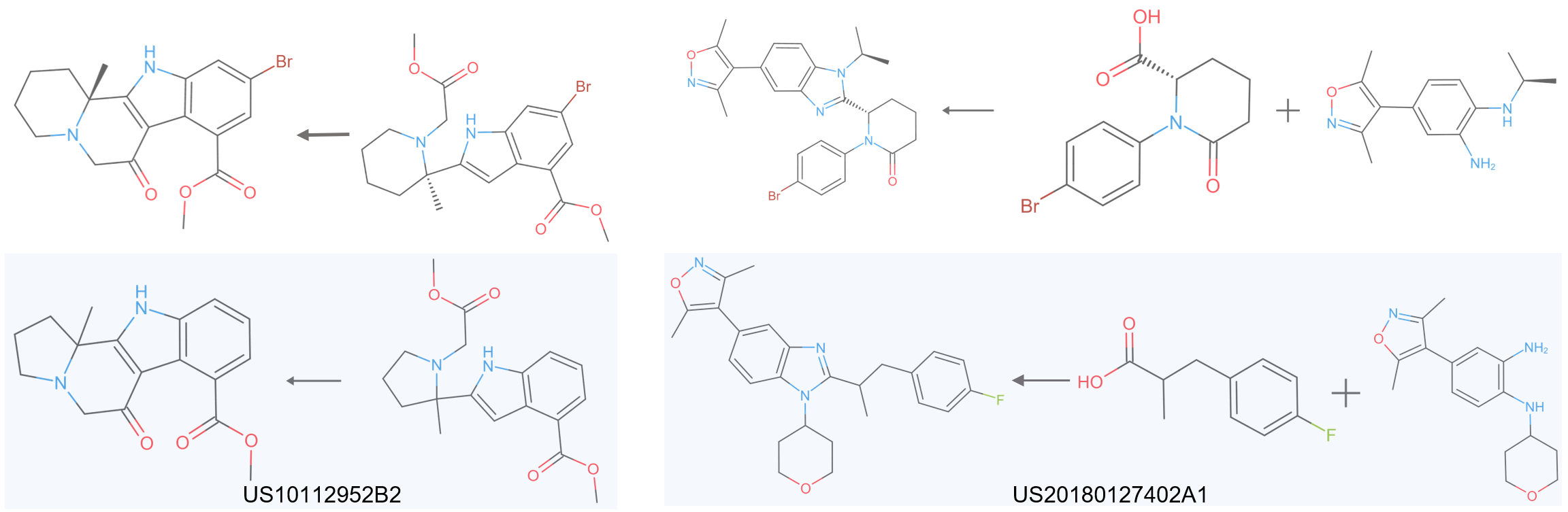}
	\caption{Selected references for proposed reactions. Blue shaded reactions are from the reaction database of U.S. patents.~\cite{us2016left,us2018right}}
	\label{figref}
\end{figure}

Besides such efficiency gains, most proposed reactions have reference reactions of satisfying resemblance that enables chemists to proceed with confidence. Showcases can be seen in \autoref{figref}, two ring-closure reactions that are supported by very similar references. 

\section{Conclusion} \label{sec:conclusion}

In this paper, we demonstrated an end-to-end, data-driven retrosynthesis system that can generate reliable synthetic routes for given molecules. During the development of ChemiRise, we provided better solutions to many existing problems such as atom mapping, one-step retrosynthesis reaction suggestion. Specifically, for the one-step retrosynthesis reaction proposer, our neural network-based model using reaction template embeddings outperforms existing models and ensures the quality of generated synthesis routes. The guiding algorithm, along with other optimizations, also contributes to the reliability and rapidity of the system by reducing unnecessary exploration of the synthesis space. The final product received mostly positive feedback from experts, in terms of overall performance and user experience. ChemiRise has been integrated into our drug discovery workflow and has provided considerable assistance to medicinal chemists in their synthesis route desings. The next steps of the demonstration of ChemiRise might include laboratory results of synthesis routes generated by our system, and a free, public version for testing users.

\bibliographystyle{unsrt}
\bibliography{references}

\begin{thebibliography}{10}

\bibitem{corey1991retrosynth}
Elias~James Corey.
\newblock The logic of chemical synthesis: multistep synthesis of complex
  carbogenic molecules (nobel lecture).
\newblock {\em Angewandte Chemie International Edition in English},
  30(5):455--465, 1991.

\bibitem{corey1971centenary}
EJ~Corey.
\newblock Centenary lecture. computer-assisted analysis of complex synthetic
  problems.
\newblock {\em Quarterly Reviews, Chemical Society}, 25(4):455--482, 1971.

\bibitem{corey1980lhasa1}
EJ~Corey, A~Peter Johnson, and Alan~K Long.
\newblock Computer-assisted synthetic analysis. techniques for efficient
  long-range retrosynthetic searches applied to the robinson annulation
  process.
\newblock {\em The Journal of Organic Chemistry}, 45(11):2051--2057, 1980.

\bibitem{corey1985lhasa2}
Elias~James Corey, Alan~K Long, and Steward~D Rubenstein.
\newblock Computer-assisted analysis in organic synthesis.
\newblock {\em Science}, 228(4698):408--418, 1985.

\bibitem{pensak1977lhasa}
David~A Pensak and Elias~James Corey.
\newblock Lhasa—logic and heuristics applied to synthetic analysis.
\newblock ACS Publications, 1977.

\bibitem{silver2017mastering}
David Silver, Julian Schrittwieser, Karen Simonyan, Ioannis Antonoglou, Aja
  Huang, Arthur Guez, Thomas Hubert, Lucas Baker, Matthew Lai, Adrian Bolton,
  et~al.
\newblock Mastering the game of go without human knowledge.
\newblock {\em nature}, 550(7676):354--359, 2017.

\bibitem{senior2020improved}
Andrew~W Senior, Richard Evans, John Jumper, James Kirkpatrick, Laurent Sifre,
  Tim Green, Chongli Qin, Augustin {\v{Z}}{\'\i}dek, Alexander~WR Nelson, Alex
  Bridgland, et~al.
\newblock Improved protein structure prediction using potentials from deep
  learning.
\newblock {\em Nature}, 577(7792):706--710, 2020.

\bibitem{klucznik2018efficient}
Tomasz Klucznik, Barbara Mikulak-Klucznik, Michael~P McCormack, Heather Lima,
  Sara Szymku{\'c}, Manishabrata Bhowmick, Karol Molga, Yubai Zhou, Lindsey
  Rickershauser, Ewa~P Gajewska, et~al.
\newblock Efficient syntheses of diverse, medicinally relevant targets planned
  by computer and executed in the laboratory.
\newblock {\em Chem}, 4(3):522--532, 2018.

\bibitem{coley2017computer}
Connor~W Coley, Luke Rogers, William~H Green, and Klavs~F Jensen.
\newblock Computer-assisted retrosynthesis based on molecular similarity.
\newblock {\em ACS central science}, 3(12):1237--1245, 2017.

\bibitem{weininger1988smiles}
David Weininger.
\newblock Smiles, a chemical language and information system. 1. introduction
  to methodology and encoding rules.
\newblock {\em Journal of chemical information and computer sciences},
  28(1):31--36, 1988.

\bibitem{smirk}
Daylight.
\newblock daylight theory: smirks - a reaction transform language.

\bibitem{tetko2020transformer}
Igor~V Tetko, Pavel Karpov, Ruud Van~Deursen, and Guillaume Godin.
\newblock State-of-the-art augmented nlp transformer models for direct and
  single-step retrosynthesis.
\newblock {\em Nature communications}, 11(1):1--11, 2020.

\bibitem{liu2017retrosynthetic}
Bowen Liu, Bharath Ramsundar, Prasad Kawthekar, Jade Shi, Joseph Gomes, Quang
  Luu~Nguyen, Stephen Ho, Jack Sloane, Paul Wender, and Vijay Pande.
\newblock Retrosynthetic reaction prediction using neural sequence-to-sequence
  models.
\newblock {\em ACS central science}, 3(10):1103--1113, 2017.

\bibitem{dai2020retrosynthesis}
Hanjun Dai, Chengtao Li, Connor~W Coley, Bo~Dai, and Le~Song.
\newblock Retrosynthesis prediction with conditional graph logic network.
\newblock {\em arXiv preprint arXiv:2001.01408}, 2020.

\bibitem{szymkuc2016computer}
Sara Szymku{\'c}, Ewa~P Gajewska, Tomasz Klucznik, Karol Molga, Piotr Dittwald,
  Micha{\l} Startek, Micha{\l} Bajczyk, and Bartosz~A Grzybowski.
\newblock Computer-assisted synthetic planning: The end of the beginning.
\newblock {\em Angewandte Chemie International Edition}, 55(20):5904--5937,
  2016.

\bibitem{segler2018planning}
Marwin~HS Segler, Mike Preuss, and Mark~P Waller.
\newblock Planning chemical syntheses with deep neural networks and symbolic
  ai.
\newblock {\em Nature}, 555(7698):604--610, 2018.

\bibitem{lowe2012extraction}
Daniel~Mark Lowe.
\newblock {\em Extraction of chemical structures and reactions from the
  literature}.
\newblock PhD thesis, University of Cambridge, 2012.

\bibitem{mo2021evaluating}
Yiming Mo, Yanfei Guan, Pritha Verma, Jiang Guo, Mike~E Fortunato, Zhaohong Lu,
  Connor~W Coley, and Klavs~F Jensen.
\newblock Evaluating and clustering retrosynthesis pathways with learned
  strategy.
\newblock {\em Chemical Science}, 12(4):1469--1478, 2021.

\bibitem{dittmar1983cas}
Paul~G Dittmar, Nick~A Farmer, William Fisanick, Reginald~C Haines, and Joseph
  Mockus.
\newblock The cas online search system. 1. general system design and selection,
  generation, and use of search screens.
\newblock {\em Journal of Chemical Information and Computer Sciences},
  23(3):93--102, 1983.

\bibitem{liu2019chemi}
Ke~Liu, Xiangyan Sun, Lei Jia, Jun Ma, Haoming Xing, Junqiu Wu, Hua Gao, Yax
  Sun, Florian Boulnois, and Jie Fan.
\newblock Chemi-net: a molecular graph convolutional network for accurate drug
  property prediction.
\newblock {\em International journal of molecular sciences}, 20(14):3389, 2019.

\bibitem{vinyals2015order}
Oriol Vinyals, Samy Bengio, and Manjunath Kudlur.
\newblock Order matters: Sequence to sequence for sets.
\newblock {\em arXiv preprint arXiv:1511.06391}, 2015.

\bibitem{us2016left}
Changyou Zhou, Bo~Ren, and Hexiang Wang.
\newblock Fused tetra or penta-cyclic dihydrodiazepinocarbazolones as parp
  inhibitors, 2011.

\bibitem{us2018right}
Neil~Anthony Pegg, David Michel~Adrien Taddei, Stuart~Thomas Onions, Eric
  Sing~Yuen Tse, Richard~James Brown, David~Kenneth Mycock, David Cousin, and
  Anil Patel.
\newblock Isoxazolyl substituted benzimidazoles, 2016.

\end{thebibliography}

\end{document}